\documentclass{iau}
\usepackage{graphicx,natbib}
 
\newcommand{\apj}{ApJ}           % Astrophysical Journal
\newcommand{\apjl}{ApJ}           % Astrophysical Journal
\newcommand{\mnras}{MNRAS}       % Monthly Notices of the RAS

\newcommand{\aap}{A\&A}
\newcommand{\araa}{ARA\&A}

\newcommand{\pasp}{PASP}
\newcommand{\apjs}{ApJS}           % Astrophysical Journal

\title{S7 : Probing the physics of Seyfert Galaxies through their ENLR \& HII Regions}

\author[Dopita et al.]{
Michael A. Dopita$^{1,2}$,
Prajval Shastri$^{3}$,
Julia Scharw\"achter$^{4}$,
Lisa J. Kewley$^{1,5}$,
Rebecca Davies$^{1}$,
Ralph Sutherland$^{1}$,
Preeti Kharb$^{3}$,
Jessy Jose$^{3}$,
Harish Bhatt$^{3}$,
S. Ramya$^{3}$,			
Elise Hampton$^{1}$,
Chichuan Jin$^{6}$,
Julie Banfield$^{7}$,
Ingyin Zaw$^{8}$,
Shweta Srivastava$^{9}$
  \and Bethan James$^{10}$
}

\affiliation{
$^{1}$RSAA, Australian National University, Cotter Road, Weston Creek, ACT 2611, Australia \\
$^{2}$Astronomy Department, King Abdulaziz University, P.O. Box 80203, Jeddah, Saudi Arabia \\
email: {\tt Michael.Dopita@anu.edu.au} \\\bigskip
$^{3}$Indian Institute of Astrophysics, Koramangala 2B Block, Bangalore 560034, India \\
$^{4}$LERMA, Observatoire de Paris, 61 Avenue de l'Observatoire, 75014 Paris, France \\
$^{5}$Institute for Astronomy, University of Hawaii, 2680 Woodlawn Drive, Honolulu, HI, USA \\
$^{6}$Department of Physics, University of Durham, South Road, Durham DH1 3LE, UK \\
$^{7}$CSIRO Astronomy \& Space Science, P.O. Box 76, Epping NSW, 1710 Australia \\
$^{8}$New York University (Abu Dhabi) , 70 Washington Sq. S, New York, NY 10012, USA \\
$^{9}$Gorakhpur University, Gorakhpur, Uttar Pradesh, India \\
$^{10}$ Institute of Astronomy, Cambridge University, Madingley Road, Cambridge CB3 0HA, UK
}

\pubyear{2014}
\volume{309}
\jname{Galaxies in 3D across the Universe}
\editors{B. L. Ziegler, F. Combes, H. Dannerbauer, M. Verdugo, eds.}

\begin{document}

\maketitle

\begin{abstract}
Here we present the first results from the \emph{Siding Spring Southern Seyfert Spectroscopic Snapshot Survey} (S7) which aims to investigate the physics of $\sim140$ radio-detected southern active Galaxies with $z<0.02$ through Integral Field Spectroscopy using the Wide Field Spectrograph (WiFeS). This instrument provides data cubes of the central $38\times25$ arc sec. of the target galaxies in the waveband $340-710$nm with the unusually high resolution of $R=7000$ in the red ($530-710$nm), and $R=3000$ in the blue ($340-560$nm). These data provide the morphology, kinematics and the excitation structure of the extended narrow-line region, probe relationships with the black hole characteristics and the host galaxy, measures host galaxy abundance gradients and the determination of nuclear abundances from the HII regions. From photoionisation modelling, we may determine the shape of the ionising spectrum of the AGN, discover whether AGN metallicities differ from nuclear abundances determined from HII regions, and probe grain destruction in the vicinity of the AGN. Here we present some preliminary results and modelling of both Seyfert galaxies observed as part of the survey.
\end{abstract}

\firstsection
\section{Introduction}
The study of nearby Seyfert galaxies offers potential insights not only into the physics of active galaxies themselves, but also into the galactic environments which feed the nuclear activity. According to the ``standard'' unified model of AGN \citep{Antonucci:90apj,Antonucci:93araa} and its extensions \citep{Dopita:97pasa}, the Seyfert~1 galaxies are seen pole-on relative to the accretion disk, and these display very broad permitted lines originating in rapidly moving gas close to the central engine. In the Seyfert~2 galaxies, the thick, dusty and toroidal accretion disk obscures the central engine, and an Extended Narrow Line Region (ENLR) often confined within an ``ionisation cone" is observed.

The nature of this ENLR can provide vital clues about the nature of the central black hole, and the mechanisms which produce the extreme UV (EUV) continuum. Seyfert galaxies are known to occupy a very restricted range of line ratios when plotted on the well-known BPT diagram \citep{Baldwin:1981ab} which plots [N II] $\lambda$6584/H$\alpha$ vs. [O III] $\lambda$5007/H$\beta$ or on the other diagrams introduced by \citet{VO87} involving the  [S II] $\lambda$6717,31/H$\alpha$ ratio or the [O I] $\lambda$6300/H$\alpha$ ratio in the place of the [N II] $\lambda$6584/H$\alpha$ ratio. It now seems clear that this is because the ENLR is (in perhaps all cases) radiation pressure dominated  \citep{Dopita:02apj,Groves:04apjs, Groves:04apj}. In this model, radiation pressure (acting upon both the gas and the dust) compresses the gas close to the ionisation front so that at high enough radiation pressure, the density close to the ionisation front scales as the radiation pressure, and the local ionisation parameter in the optically-emitting ENLR becomes constant. This results in an ENLR spectrum which is virtually independent of the input ionisation parameter. For dusty ENLR the radiation pressure comes to dominate the gas pressure for $\log U \gtrsim -2.5$, and the optical emission spectrum becomes invariant with the input ionisation parameter for $\log U \gtrsim 0.0$. In this condition, the observed density in the ENLR should drop off in radius in lockstep with the local intensity of the radiation field; $n_e \propto r^{-2}$, and the EUV luminosity can be inferred directly from a knowledge of the density and the radial distance.

Despite this constancy of the emission spectrum at high ionisation parameter, the emission line ratios remain sensitive to the form of the input EUV spectrum. Although the spectral energy distribution of Seyfert galaxies has now been studied from the far-infrared all the way up to hard X-ray and even $\gamma-$ray energies (\emph{e.g.} \citet{Done:12mnras,Jin:12a,Jin:12b}), the spectral region between 13.6 and $\sim 150$eV remains inacessible to either ground- or space-based observation. However, the ENLR spectrum is most sensitive to the form of this EUV spectrum, and its gross features can be inferred by a method reminiscent of the energy balance or Stoy technique used to estimate the effective temperature of stars in planetary nebulae \citep{Stoy1933,Kaler1976,Preite-Martinez1983}. As the radiation field becomes harder, the heating per photoionisation increases, and the sum of the fluxes of the forbidden lines becomes greater relative to the recombination lines. Furthermore, individual line ratios are sensitive in different ways to the form of the EUV spectrum, and this can be exploited to infer the form of the EUV spectrum. However, for this method to work, we need to have a well-constrained knowledge of the chemical abundances in the ENLR.

A fair number of high-resolution imaging surveys of Seyferts in the [O III] $\lambda 5007$ line have been undertaken, notably by \citet{Pogge88a, Pogge88b, Pogge89, Haniff88, Mulchaey96a,  Mulchaey96b, Falcke98} and \citet{Schmitt03}. However, up to the present there have been relatively few systematic spectroscopic studies of the ENLR of nearby Seyferts at optical wavelengths \citep{Cracco11}.  Notable amongst these is the multi-slit work by \citet{Allen:99apj}. However, the advent of integral field spectroscopy has made such studies much more efficient, and it is now possible to gather as much data in a night as \citet{Allen:99apj} did in three years! Motivated by these ideas, we have undertaken the \emph{Siding Spring Southern Seyfert Spectroscopic Snapshot Survey} (S7) and here describe the methodology and first results of this survey.

\section{The S7 Survey}
The S7 survey is an integral field survey in the optical of $\sim 140$ southern Seyfert and LINER galaxies. It uses the Wide Field Spectrograph (WiFeS) mounted on the Nasmyth focus of the ANU 2.3m telescope \citep{Dopita:2010aa}. This instrument provides data cubes of the central $38\times25$ arc sec. at a spatial resolution of 1.0 arc sec. It covers the waveband $340-710$nm with the unusually high resolution of $R=7000$ in the red ($530-710$nm), and $R=3000$ in the blue ($340-560$nm). The typical throughput of the instrument (top of the atmosphere to back of the detector) is $20-35$\% \citep{Dopita:2010aa}, which provides an excellent sensitivity to faint low surface brightness ENLR features.

The sample objects for S7 were selected from the \citet{VC06} catalogue of active galaxies, which is the most comprehensive compilation of active galaxies in the literature. Since we wish to investigate the interaction of the bipolar plasma jets with the Narrow Line Region and the ISM of the host galaxy, we limited the sample to galaxies with radio flux densities high enough to permit 
radio aperture synthesis observations.  We adopted the following selection criteria:
\begin{itemize}
\item Declination  $<10$ degrees to avoid WiFeS observations at too great a zenith distance,
\item Galactic latitude $<|20|$ degrees (with a few exceptions) to avoid excessive galactic extinction.
\item Radio flux density at 20cm $\lesssim20mJy$ for those targets with declination N of -40 deg, which have NVSS measurements, and
\item Redshift $<0.02$. This criterion ($D< 80$ Mpc) ensures that the spatial resolution of the data is better than 400 pc arcsec$^{-1}$, sufficient to resolve the ENLR, and to ensure that the important  diagnostic [S~II] lines are still within the spectral range of the WiFeS high-resolution red grating.
\end{itemize}

All the data are reduced using the using the {\tt PyWiFeS} pipeline written for the instrument \citep{Childress14}. In brief, this produces a data cube which has been wavelength calibrated, sensitivity corrected (including telluric corrections), photometrically calibrated, and from which the cosmic ray events have been removed. With this software, a full night's worth of data can be reduced on a laptop in about 2 Hr. The data cube is then passed through the integral field spectrograph toolkit {\tt LZIFU}  (Ho et al. 2014 in prep.) to extract gas and stellar kinematics and other parameters from the WiFeS data. {\tt LZIFU} performs simple stellar population (SSP) synthesis fitting to model the continuum, and fits the emission lines as 3-component Gaussians.

\begin{figure}
\centering
\includegraphics[width=0.75\columnwidth]{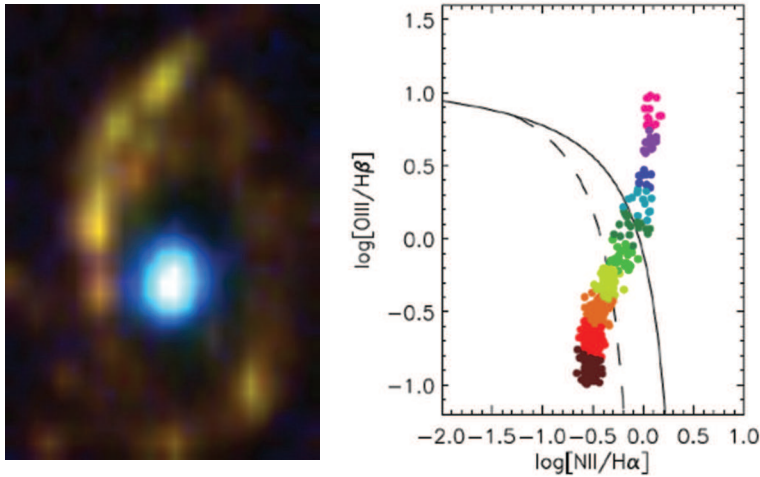} 
\caption{ An example of data products produced in the S7 survey. The Left-hand panel is an image of NGC~6890 in H$\alpha$ (red), [N II] $\lambda 6584$ (green) and [O III] $\lambda 5007$ (blue), boxcar smoothed over one arc sec. The H II regions appear reddish or gold depending on their excitation, and trace out the spiral arms, while the nucleus and the ENLR appears blue. On the right-hand panel we show the BPT diagram for the individual spaniels, colour coded according to the angular distance from the nucleus. This forms a tight mixing sequence between pure AGN line ratios (magenta) and pure HII region emission (brown). Compare \emph{e.g.} \citet{Dopita:14aa, Davies14a}. }\label{fig:fig1}
\end{figure}

The standard data products of the S7 survey include images and BPT diagrams such as shown for one galaxy, NGC~6890 in Figure \ref{fig:fig1}, and nuclear spectra -- examples of which are given in Figure \ref{fig:fig2}. In addition we have (for the gaseous component) emission line flux maps, line ratio maps, gas velocity maps, and gas velocity dispersion maps. From these we can derive abundance, star formation and dust extinction maps. For the stellar components we have continuum maps, extinction, stellar velocity and stellar velocity dipsersion maps. From these we can derive age distribution, metallicity etc.

\section{First Results}
In the first data S7 data release of about 50 Seyferts and LINERS (due before the end of 2014), a wide range of properties will be explored. Here we have only space to point out a couple of highlights of the research accomplished so far.
\subsection{Form of the EUV spectrum}
As described above, the ENLR spectrum provides strong constraints on the EUV spectrum of the Seyfert nucleus, provided that the ionisation parameter and chemical abundances in the ENLR can be inferred. As a first test example of the technique \citet{Dopita:14aa} examined NGC~5427 in some detail. Using off-nuclear integrations as well as the on-nucleus field, we were able to establish the abundance gradient and determine nuclear abundances from strong emission lines of individual HII regions using the {\tt Pyqz} photoionisation grid interpolation routine described in \citet{Dopita13a}. With the inferred nuclear abundances, a tight constraint on both $\log U$ and the shape of the EUV spectraum was obtained using the same {\tt Mappings IV} code that was employed for the HII region analysis. This demonstrated that the hot accretion disk + hard Compton model of \citet{Done:12mnras,Jin:12a,Jin:12b} worked well. The intermediate Compton component introduced by these authors in order to fit the X-ray data of luminous Seyferts seems to be absent. However, our models seem to indicate that such a component arises naturally within the inner ENLR itself, provided that the ionisation parameter is sufficient to Compton heat the plasma up to $\sim10^6$K.

\subsection{Coronal Emission Regions}
\begin{figure}
\centering
\includegraphics[width=0.75\columnwidth]{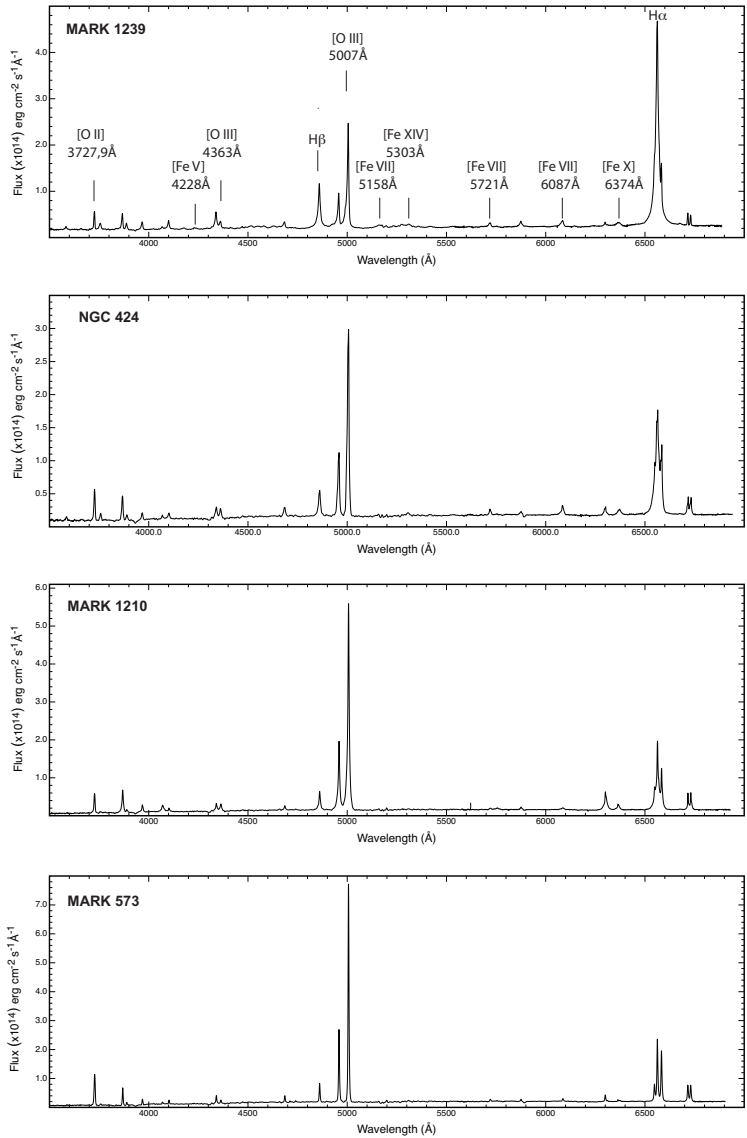} 
\caption{A selection of S7 galaxy nuclear spectra showing strong coronal line emission. The main coronal species and a number of the other lines are identified on the first panel. The panels are ordered in terms of the electron density as indicated by the [O III] $\lambda \lambda 4363/5007$\AA\ ratio (MARK 1239 being the densest). Note that this order matches the order of the H$\alpha$ line width and the [N II] $\lambda 6584$/H$\alpha$ ratio. This suggests strong correlations between the region of the narrow line emission, the electron density, and the strength of the coronal features. }\label{fig:fig2}
\end{figure}

Nuclear spectra of very high quality have been extracted for all the observed S7 galaxies using a 2.0 arc sec. aperture. Although the analysis of these is not yet complete, it is already clear that not only are there a number of mis-classified galaxies in the \citet{VC06} catalogue, but also the that classification scheme used for Seyferts (S3b, S1.8, S3h etc.) is too fine, and perhaps undermines a sense of unity and continuity between the different classes of object. It is clear from our data that important systematic trends exist object to object, and here we seek to illustrate this within the restricted range of objects showing clear coronal emission from species such as [Fe V], [Fe VII], [Fe X] and [Fe XIV], see Figure \ref{fig:fig2}. The objects shown here are not all the objects for which we have detected coronal emission -- strong coronal emission can also be seen in the Type I objects, Fairall 51 and NGC~7469. It is evident from Figure \ref{fig:fig2} that the relative strength and excitation of the coronal emission is correlated with the H$\alpha$ line width, and with the electron density in the [O III] - emitting region, as evidenced by the [O III] $\lambda\lambda 4363/5007$ ratio. The density in the low-excitation gas as revealed by the [S II] $\lambda\lambda 6717/6731$ ratio remains around $n_e \sim 10^4$cm$^{-3}$, and the strength of these [S II] lines and the [N II] $\lambda 6584$ line relative to the broad component of H$\alpha$ also changes systematically with H$\alpha$ line width. This strongly suggests that these low excitation lines arise in a region which is physically distinct from the region emitting the coronal species.

These properties are consistent with the model advocated by \citet{Mullaney09}. In this, the coronal lines arise in a dense gas launched from the dusty inner torus ($10^{17} < R/{\rm cm} < 10^{18}$) at very high local ionisation parameter $\log U \sim -0.4$, and is accelerated by radiation pressure to a terminal velocity of a few hundred km~s$^{-1}$. At this ionisation parameter the gas is Compton heated to $\sim 10^6$K or greater and the dust in the coronal emission region is destroyed, allowing the forbidden iron lines to reach such high intensity relative to the hydrogen lines. Systematic changes in width, excitation and and density are consistent with different inner torus radii for the objects shown in Figure \ref{fig:fig2} - MARK~1239 having a small inner torus, and MARK~573 having a larger one. 

\section*{Acknowledgements}
\noindent M.D. and L.K. acknowledge the support of the Australian Research Council (ARC) through Discovery project DP130103925. M.D. also acknowledges support under the King Abdulaziz University {\em HiCi} program. This research has made use of the NASA/IPAC Extragalactic Database (NED), which is operated by the Jet Propulsion Laboratory, California Institute of Technology, under contract with the National Aeronautics and Space Administration, the NASA Astrophysics Data System (ADS), and SAOImage DS9 (Joye \& Mandel 2003), developed by the Smithsonian Astrophysical Observatory.

\end{document}